\documentclass[11pt]{article}
\newcommand{\be}{\begin{equation}}
\newcommand{\ee}{\end{equation}}
\usepackage{amsmath}
\usepackage{amsfonts}
\usepackage{amssymb}
\usepackage[colorlinks=true, urlcolor=blue, linkcolor=green]{hyperref}
\usepackage{float}

\usepackage{graphicx}
\begin{document}
	
	\begin{center}
		{\Large{\textbf{Morphology and high frequency bio-electric fields}}}\\ 
		Johann Summhammer\\ 
		{\small{Vienna University of Technology,
				Institute of Atomic and Subatomic Physics\\ 
				Stadionallee 2, 1020 Vienna, Austria\\
			    %Started: 15 March 2021\\
		}}
	\end{center}

\begin{abstract}
	We investigate possible shapes of the electric field, which oscillating dipoles in a certain region of biological tissue can produce in a neighboring region, or outside the tissue boundaries. We find that a wide range of shapes, including the typical morphology of limbs and appendages, can be generated as a zone of extremely low field amplitudes embedded in a zone of much larger field amplitudes. Neutral molecules with a resonance close to the frequency of the oscillating field may be attracted to this zone or be repelled from it, while the driving effect on molecules with an electric charge is only extremely weak. The forces would be sufficient for the controlled deposition of molecules during growth or regeneration. They could also serve as a method of information transfer. 
\end{abstract}

\section{Introduction}
In living organisms, biomolecules of a wide range of sizes must be transported to various locations where they are needed for specific reactions. The mechanisms driving transport can be based on simple physical principles like Brownian motion, possibly constrained by intracellular or extracellular structures \cite{Hoefling2013}, or on static and slowly varying electric fields \cite{DeLoof1985, Cunningham2012, Funk2015}. For higher functions within the cell organized schemes come into play, for instance, the transport of proteins and organelles along microtubules \cite{Burute2019, Mogre2020}. The static or slowly varying endogenous electric fields seem to be particularly important during embryogenesis \cite{Nuccitelli2003} and during regrowth of amputated or shed body parts in animals capable of such regeneration \cite{Durant2019, Durant2017, Mitoh2021}. They are also instrumental in the similar function of wound healing, where they are found universally from plants to humans \cite{Tyler2017}. Aside from such quasi-static electric fields, the energetic processes in living systems are often unavoidably accompanied by temporary oscillating electric fields. In the optical domain (infrared to ultraviolet, 10$^{15}$ -10$^{16}$ Hz), emissions in the form of bioluminescence have been known ever since, e.g. fireflies, while weak external and internal emissions, termed biophotons \cite{Chang1998}, have come into focus only more recently \cite{Zapata2021, Kumar2016, Tang2014}. The functional role of these optical phenomena is due to the energy of such photons, which is sufficient for chemical reactions both within an organism as well as in remote organisms, where they can be detected with various forms of eyes or simpler optical sensors. However, the present paper is more concerned with the frequency range below the infrared, starting at wavelengths from some 10 micrometers (10 THz) up to several centimeters (10 GHz) and perhaps meters (100 MHz). This is the frequency range of telecommunication and extensive work has been done on the possible health risks of these electromagnetic oscillations (e.g. \cite{Romanenko2017}). But there are only few reliable measurements on emissions of biological systems in this frequency range. For instance, yeast cells emit at 42 GHz during the M-phase of cell mitosis \cite{Jelinek2007}. The significance of such emissions is unclear, since so far no purposeful functions seem to have been identified \cite{Cucera2015}. In a way this is surprising, because this is the range of natural vibrations in biomolecules \cite{Powell1987}-\cite{Qin2019}. 
Many of these molecules, or parts of these molecules, have an electric dipole moment, and therefore could be emitters as well as receivers of such radiation. If the intensity of the radiation is above the thermal background it should have some effect, which evolution should have integrated either as a negligible disturbance or as a useful resource. In fact, it has been hypothesized by Fröhlich quite some time ago, that the natural emission and absorption frequencies of biomolecules should lead to resonant interaction between molecules of the same kind and much more energy should be condensed into such modes than attributable from thermal considerations alone \cite{Froehlich1968}-\cite{Froehlich1983}. Experimental evidence for proteins has been found only recently \cite{Lundholm2015}. Fröhlich's work still stimulates research on how living systems might make use of oscillatory resonances between molecules. E.g., in \cite{Olmi2018} the emergence of a collective oscillating mode of dipolar molecules driven by a random energy supply was shown, and in \cite{Tuszynski2015} it was analyzed how resonant interaction might help guiding biomolecules towards each other before a lock-and-key reaction can set in. From a physical perspective it is especially in connection with molecular transport as in embryogenisis, regeneration and wound healing, that oscillating electromagnetic fields deserve a deeper look. For, while Earnshaw's theorem limits the shapes of \emph{static} electric fields and hence the possible directions of force to relatively simple forms, the forces from \emph{oscillating} electric fields are not subject to this limitation (e.g. \cite{Earnshaw}).

In this context the present paper wants to point out a further collective effect of oscillating dipoles, which could be of relevance to processes in which molecules must be steered to specific locations, often many wavelengths away from the oscillators. It makes use of the intricate interference patterns, which can be created by an ensemble of randomly placed dipoles oscillating with the same frequency but with different amplitudes and different phases. Such an interference pattern may contain almost arbitrarily shaped zones of vanishing field amplitude, which are surrounded by regions with field amplitudes orders of magnitude larger, so that molecules with fitting resonance frequencies can be attracted to such zones. The concept for generating the oscillating field is analogous to creating a hologram, except that in the present case the desired object must appear dark rather than bright.

\section{Derivation of the low field zone}

For the purpose of illustration we imagine a salamander, whose right hand has been amputated and is about to regrow (Fig.1). The right arm and thorax are assumed to contain many oscillating dipoles whose collective electric field creates an outline of the current growth zone as a volume of extremely low field amplitude. Molecules with the appropriate resonance will be attracted to this zone, because they are driven away from the high field regions. Immediately after the amputation, the growth zone will be the surface area of the wound. At later stages it will be the surface of the new tissue, which increasingly starts to resemble the outline of the hand. But it is also possible that growth is needed in regions within the newly formed tissue, and then it is conceivable that a zone of extremely low electric field amplitude is generated there. Since the molecules to be steered to certain locations will be confined mostly within cell boundaries, the drift forces to be derived below will work within such confines, but may as such also control in which direction a cell begins to bulge. At any rate, it is not claimed here that limb regeneration works by means of oscillating electric fields. But the example of regrowth of an amputated hand is a concrete image to which we can pin the mathematical derivation, and it may underline the capabilities of bio molecular organization of such fields.

We have $N$ transmitting dipoles, all of which oscillate with the same frequency $\nu$, but with different amplitudes and with arbitrary initial phases. These dipoles are located at random positions in a source region $S$, as shown in Fig.1. The region $S$ has some extension in the x- and y-directions, and a smaller one in the z-direction, which is not shown in Fig.1. And also the target region $T$ shall be relatively thin in the z-direction. Finally, we assume that all dipoles are oriented and oscillate along the z-direction. These assumptions will allow us to consider only the z-component of the radiated electric field of the dipoles, because the component perpendicular to z will be very small. Therefore, when we refer to the electric field in the following paragraphs, we shall mean only the z-component of the electric field.
 
Let the position of dipole $j$ in the source region $S$ be denoted by the vector $\bold{R}_j$. The z-component of the radiated electric field of this dipole at a point $\bold{r}_i$ can be written as
\begin{equation}
	E_j(\bold{r}_i,t) = A_j  \frac{\sin^2\theta_{ij}} {\left|\bold{r}_i-\bold{R}_j\right|}\cos\left[2\pi\nu\left(t-\frac{\left|\bold{r}_i-\bold{R}_j\right|}{c}\right)-\varphi_j\right],
\end{equation}
where $c$ is the velocity of light (which may have to be corrected for passage of the waves through different media), $A_j$ is the amplitude factor of the electric field, $\theta_{ij}$ is the angle between the z-direction and the vector $(\bold{r}_i-\bold{R}_j)$, and $\varphi_j$ is the arbitrary but fixed phase of this oscillating dipole. Here we have assumed that the distance $\left|\bold{r}_i - \bold{R}_j\right|$ is at least several wavelengths $\lambda=c/\nu$, so that the far-field approximation is valid (e.g. \cite{farfieldwiki}). However, this is not a necessary assumption, because the derivation would be valid for any distance-dependence of the oscillating electric field.
\begin{figure}[H]
	\centering
	\includegraphics[scale=0.56]{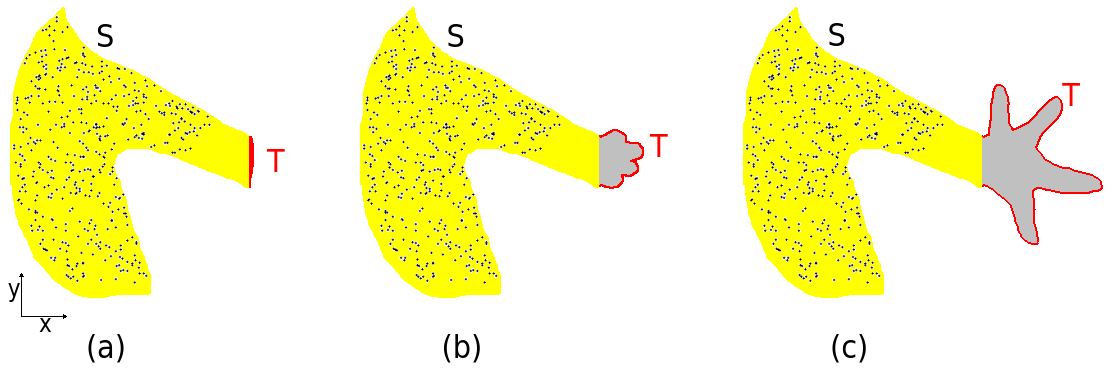}
	\caption{Different stages of regeneration of an amputated hand of a salamander. The yellow region S represents right arm and thorax and contains many randomly distributed source points (dark dots) which emit high frequency electromagnetic waves. The red outlines are the target region T, to which molecules must be transported because it is the growth region at that moment. (a) Right after amputation, (b) At a time when the outlines of the new hand become apparent, (c) End stage of regeneration. The grey areas in (b) and (c) are newly formed tissue, which is assumed as not yet containing active source points of the electromagnetic field.}
	\label{}
\end{figure}

\noindent The total electric field at point $\bold{r}_i$ is the sum of the contributions from \emph{all} oscillating dipoles:
\begin{equation}
E_{tot}(\bold{r}_i,t) = \sum_{j=1}^{N} \Big\{ A_j  \frac{\sin^2\theta_{ij}} {\left|\bold{r}_i-\bold{R}_j\right|}\cos\left[2\pi\nu\left(t-\frac{\left|\bold{r}_i-\bold{R}_j\right|}{c}\right)-\varphi_j\right] \Big\}.
\end{equation}
This equation contains the $N$ amplitude factors $A_j$ $(j=1,...,N$) as free parameters, which will have to be determined by constraints set at the target points of the low field zone. If we are only interested in the relative strength of the electric fields of different points, we can assign a value to one amplitude factor, and we choose 
\begin{equation}
A_N \equiv 1.
\end{equation} 
This leaves $N-1$ free parameters. We shall determine their values by demanding a quantity $S$, which we define as the time average of the sum of the squared electric fields at $N-1$ points $\bold{r}_i$ in the target zone, to attain a minimum:
\begin{equation}
S=\Big< \sum_{i=1}^{N-1} E_{tot}^2(\bold{r}_i,t) \Big> \rightarrow \text{Minimum}.
\end{equation}
By inserting from eq.(2) and doing the time integration over one oscillation period, $S$ becomes
\begin{equation}
S=\frac{1}{2}\sum_{i=1}^{N-1}\sum_{j=1}^{N}\sum_{k=1}^{N} A_j A_k f_{ij}f_{ik}\cos[2\pi\nu\left(\tau_{ij}-\tau_{ik}\right)-\varphi_j+\varphi_k],
\end{equation} 
where we have used the abbreviations
\begin{equation}
f_{ij}=\frac{ \sin^2 \theta_{ij} }{\left|\bold{r}_i-\bold{R}_j\right|}
\end{equation}
and
\begin{equation}
\tau_{ij}=\frac{\left|\bold{r}_i-\bold{R}_j\right|}{c}.
\end{equation}
The requirement for a minimum leads to \footnote{In principle this is also the requirement for a maximum, but $S$ has no upper bound under variations of the amplitude factors.}
\begin{equation}
\frac{\partial S}{\partial A_l}=0, \hspace{4mm}\text{for}\hspace{2mm} l=1,...,N-1.
\end{equation} 
The resulting set of $N-1$ equations can be written in the form

\begin{equation}
\begin{pmatrix}
w_{1,1} & w_{1,2} & \cdots & w_{1,N-1} \\
w_{2,1} & w_{2,2} & \cdots & w_{2,N-1} \\
\vdots & \vdots & \ddots & \vdots \\
w_{N-1,1} & w_{N-1,2} & \cdots & w_{N-1,N-1} \\
\end{pmatrix}
\begin{pmatrix}
A_1 \\ A_2 \\ \vdots \\ A_{N-1}
\end{pmatrix}
= -
\begin{pmatrix}
w_{1,N} \\ w_{2,N} \\ \vdots \\w_{N-1,N}
\end{pmatrix}.
\end{equation}
The matrix elements are obtained as
\begin{equation}
w_{l,j}=\sum_{i=1}^{N-1} f_{ij} f_{il} \cos\left[2\pi\nu\left(\tau_{ij}-\tau_{il}\right)-\varphi_j+\varphi_l\right].
\end{equation}
Solving these equations yields the amplitudes $A_1$,..., $A_{N-1}$. With $A_N$ from eq.(3) the electric field of eq.(2) can be calculated and is obtained in the general form
\begin{equation}
E_{tot}(\bold{r},t)=G(\bold{r})\cos\left[2\pi\nu t +\vartheta(\bold{r})\right].
\end{equation}
\noindent It is a harmonic oscillation at any point $\bold{r}$, (except at the source points $\bold{R}_j$ where it is undefined), and is characterized by the amplitude function $G(\bold{r})$ and the phase function $\vartheta(\bold{r})$. The latter is conveniently defined such that $G(\bold{r})$ is nowhere negative. Figure 2 shows what the amplitude function $G(\bold{r})$ might look like for the three different stages of limb regeneration depicted in figure 1. It should also be noted that $G(\bold{r})$ has very interesting properties if the distances between neighboring target points are smaller than the wavelength $\lambda$, as is in fact the case for figure 2: Then, $G(\bold{r})$ does not only vanish at the target points, but is also very small in the space between. At very large distances from the source region, $G(\bold{r})$ approaches the expected behavior of appearing to be emitted from a single point source.

Two alternative methods of deriving a low field zone are presented in appendix A, and examples of how detailed the low field zone can appear with different wavelengths are shown in appendix B.
 
\begin{figure}[H]
	\centering
	\includegraphics[scale=0.90]{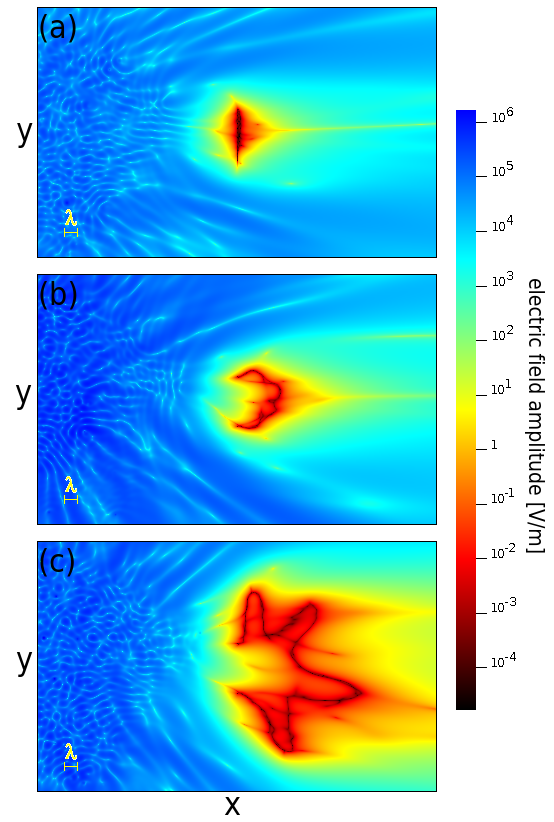}     %%{FieldAmpl-xy-xz-16.667Lambda-BestMethod.png}
	\caption{Amplitude function $G(\bold{r})$ in the x-y plane around the source and target regions for the three different stages (a)-(c) of regeneration of Fig.1. Logarithmic color scale, and expressed in real electric field values for a total emitted power of 2 mW. In each image the source region of arm and thorax appears as the finely structured dark blue region on the left, where the electric field amplitude is up to $10^{11}$ times larger than in the target region. The smallest values of the electric field amplitude are in the respective growth regions (dark red to black). The wavelength $\lambda$ of the electromagnetic oscillations is indicated in the lower left corners. The source points were distributed randomly in the x-y plane as shown in Fig.1, but also had randomly distributed z-coordinates between -4$\lambda$ and +4$\lambda$. The target points were only in the x-y plane (the target points are shown as red lines in Fig.1). }
	\label{}
\end{figure}

\section{Forces guiding molecules to the low field zone}

\subsection{Force on neutral but polarizable molecules}

First we look at molecules which carry no net electric charge, and no permanent electric dipole moment, but which exhibit an electric polarizability $\alpha$. In an oscillating electric field the induced electric dipole of such a molecule tries to align itself with the electric field to minimize its potential energy. If the electric field has a spatial gradient, this results in a force on the molecule. This force has been exploited for quite some time to control the motion of molecules and very small particles by optical tweezers \cite{Ashkin1986}, and also for optical trapping of atoms and molecules (for a review see \cite{Grimm2000}). In an electric field $\bold{E}$ the dipole moment of the molecule is
\begin{equation}
\bold{p}=\alpha \bold{E},
\end{equation}
and its time averaged potential energy in the field is
\begin{equation}
U=-\frac{1}{2}\big{<} \bold{pE} \big{>} =Re(\alpha)\tilde{E}^2,
\end{equation}
where $\tilde{E}$ is the amplitude of the oscillating electric field and $Re(\alpha)$ is the real part of the polarizability. (The imaginary part covers absorption and re-emission of energy by the induced dipole.) The force on the molecule is given by the gradient of the potential,
\begin{equation}
\bold{F}=-\bold{\nabla}U.
\end{equation}
The force also depends on the frequency, because the polarizability is frequency-dependent. Although large molecules will require a more detailed analysis, the principle behavior will be the same as that for an atom which is driven by an electric field with angular frequency $\omega$. If that frequency is not too far from the resonance frequency $\omega_0$ of an electronic excitation, and if the electric field amplitudes are sufficiently small so as not to drive the atom into a saturation \cite{Grimm2000}, the real part of the polarizability can be approximated by 
\begin{equation}
Re(\alpha)=\frac{3\pi c^3 \epsilon_0}{\omega_0^3}\frac{\Gamma}{\Delta},
\end{equation}
where
\begin{equation}
\Delta=\omega_0-\omega.
\end{equation}
Here, $\Gamma$ is a damping constant accounting for re-emission of energy by the driven atom and $\epsilon_0$ is the dielectric constant of vacuum. For large molecules this factor will certainly be different, and in a medium moreover the index of refraction and the appropriate dielectric constant will have to be included. But the strong increase of the polarizability with a decrease of the detuning $\Delta$ will be the same, as well as the change of sign when the driving frequency goes from below to above the resonance frequency (red or blue detuning, respectively). This change of sign means that the force on the atom or molecule can be directed towards regions of greater or towards regions of smaller electric field amplitudes. In our case the molecules shall be directed towards the zone of weakest electric field, and therefore the frequency of the oscillating electric field should be blue-detuned relative to the resonance frequency of the molecule, therefore $\omega > \omega_0$ or
\begin{equation}
\Delta < 0.
\end{equation}

It will now be interesting to estimate the amount of power which a salamander could possibly emit in the form of a high frequency electric field, and which forces on biomolecules could possibly result from this. As an example we take a red spotted newt with a weight of 4 grams. At 25°C its metabolic oxygen rate is around 138$\pm$30 $\mu$l O$_2$/g/hr at rest, and around 275$\pm$87 $\mu$l O$_2$/g/hr at moderate activity \cite{Jiang1993}. This rate would be much lower at lower temperature, because newts are cold-blooded animals, but it is known that newts in the process of regenerating a lost forelimb prefer to stay in places with ambient temperatures between 24° and 25°C \cite{Tattersall2012}. For the metabolic energy conversion we will assume the human value of 4.7 kcal per liter of oxygen \cite{hyperphysics}, because metabolic processes are very similar in all species. Assuming the metabolic oxygen rate of 275 $\mu$l O$_2$/g/hr, our newt of 4 grams will expend about 6 mW of power at 25°C. If it devotes one third to the regeneration of the hand, the oscillating dipoles in arm and thorax of Figs. 1 and 2 could be considered to emit around 2 mW.

This value has been used to scale the electric field amplitudes mapped in Fig.2. To this end the  whole source region S has been represented by a single oscillating dipole with maximum dipole moment $p_0$ and the electric field at a very large distance has been calculated. The time averaged power of such a dipole is given by
\begin{equation}
\bar{P}=\frac{n^3\omega^4 p_0^2}{12\pi\epsilon_0\epsilon c^3},
\end{equation}
where $n$ and $\epsilon$ are the medium's index of refraction and relative dielectric permittivity, respectively. The electric field amplitude in the plane of the oscillating dipole and perpendicular to it, at a distance $r$ very far from the dipole, can be written as
\begin{equation}
\tilde{E}_\perp \approx \frac{n^2 \omega^2}{4\pi\epsilon_0\epsilon c^2}\frac{p_0}{r}
\end{equation}
and by expressing $p_0$ through $\bar{P}$ from eq.(18) it becomes \cite{RadiatingDipole}
\begin{equation}
\tilde{E}_\perp \approx \frac{n}{r} \sqrt{\frac{3 \bar{P}}{4\pi\epsilon_0\epsilon c} } .
\end{equation}

In order to connect the amplitude function $G(r)$ shown in Fig.2 to  the real electric field amplitude at the assumed power emission of 2 mW, the value of $G$ was determined at several points at a distance of 1000$\lambda$ from the emission center, and the corresponding electric field amplitude $\tilde{E}_\perp$ was calculated by use of eq.(20). For the index of refraction and for the dielectric constant of the medium the values for water around 1 THz were chosen (e.g.\cite{Liu2019}): $n$=2.1 and $\epsilon$=4. The length scale was set by assuming the distance between the tips of first and last finger in Fig.2c to be 5 mm. These numbers also determined the wavelength of the electromagnetic oscillations in the medium as $\lambda$=0.366 mm and the frequency as $\nu$=390 GHz. With these data it is now possible to extract numbers for the $\emph{gradient}$ of the electric field amplitude from Fig.2. For instance, we find many instances close to the growth zone where the color changes from green (100 V/m) to orange (1 V/m) within a distance of around one wavelength. This translates to a gradient of around 2.7$\times$10$^5$ V/m$^2$. If we now assume the polarizablity of the biomolecule of interest, at a frequency close to a resonance, to be such as to result in an effective dipole of 1000 electron charges times a distance of 1 nm - which may be possible because even small biomolecules influence hundreds of water molecules around \cite{Leitner2008} - we obtain a drift force on the biomolecule of around 4.33$\times$10$^{-20}$ Newton. This molecule drifts in the aqueous environment of the healing wound and will be exposed to a friction force according to Stokes' law
\begin{equation}
F_{fric} = 6\pi \mu R v,
\end{equation}
where $\mu$ is the viscosity of water (8.9$\times$10$^{-4}$ Pa$\cdot$s at 25°C), $R$ is the radius of the biomolecule which we assume as 1 nm, and $v$ is the drift velocity to be determined. $F_{fric}$ and the driving force will cancel each other, and we obtain $v\approx 4.18$ nm/s. In the course of a day this comes to a distance of 0.36 mm. While there is quite some uncertainty in this estimate, e.g., diffusion in organic water mixtures can be faster than expected by eq.(21) \cite{Evoy2020}, or the electric field gradient could be considerably larger if we assumed the source points of the electromagnetic oscillations to be closer to the growth regions than assumed here (see Fig.1), it lies in the vicinity of what would be needed, considering that regeneration of a forelimb takes 45 days or more \cite{Joven2019}.

\subsection{Force on molecules with a net charge}

A molecule with a net electric charge in an oscillating electric field will experience a periodic force which drives it up and down along the field direction. If the electric field has a spatial gradient along the same direction, the distances covered in successive half cycles of the electric field will no longer cancel each other and there will be a net motion of the molecule towards the region of weaker field, independent of the sign of the charge. The effective driving force, called the ponderomotive force, can easily be derived for a standing wave electric field, but has no general solution for our case of traveling waves \cite{Burton2017}. Nevertheless, a simple estimate of the drift velocity of a charged molecule can be made. Let the molecule have an effective charge $Q$, a mass $m$ and a radius $R$. The oscillating field shall be parallel to the z axis, and the gradient of the field shall also be along z. 
We are interested in the distance the molecule moves during one full oscillation of the electric field. Initially the molecule shall be at rest at the origin and the electric field amplitude there shall be $E_0$. The gradient of the electric field shall be $\nabla E$. Thus the force on the molecule is initially given by
\begin{equation}
F(t)= QE_0\cos(2\pi\nu t)
\end{equation}
We shall assume that the molecule moves very little during one half cycle of the oscillation, so that the amplitude $E_0$ is valid throughout the first half cycle. We shall also assume that the force from the electric field shall be at equilibrium with the friction force $F_{fric}$ in water (eq.(21)) at any time. This implies that the velocity shall always be proportional to the electric field. Combining equations (21) and (22) gives the velocity of the molecule during the first half cycle,
\begin{equation}
v(t)= \frac{QE_0}{6\pi\mu  R}\cos(2\pi\nu t).			
\end{equation}
By integration we obtain the distance traveled during this time,
\begin{equation}
z_{1/2}=\frac{QE_0}{6\pi^2\mu  R \nu }.
\end{equation}
At the position $z_{1/2}$ the electric field amplitude is slightly different due to the spatial gradient $\nabla E$, and therefore the force during the second half cycle will be
\begin{equation}
F(t)= Q(E_0+z_{1/2}\nabla E)\cos(2\pi\nu t).
\end{equation}
This leads to a changed velocity and therefore to a changed traveled distance during the second half cycle, and the molecule will not return to its original position. Its effective displacement after one full cycle turns out to be
\begin{equation}
z_1=-\frac{Q^2E_0 \nabla E}{\left(6\pi^2\mu  R \nu \right)^2 },
\end{equation}
and the corresponding effective velocity is
\begin{equation}
v_1=z_1 \nu = -\frac{Q^2E_0 \nabla E}{\left(6\pi^2\mu  R \right)^2 \nu}.
\end{equation}
To obtain an idea of how large $v_1$ is, we take the parameters as in the previous section and assume an exaggerated value of 100 elementary charges for the charge on the molecule \cite{Zhang2019}. We find $v_1 \approx 2.5 \times 10^{-17}$ m/s. This is extremely small, so that the ponderomotive force from very high frequency electric fields will not play a role for the migration of charged biomolecules.

\section{Discussion}

We have seen in the derivations of section 2 (and of the appendix) that collective oscillations of biomolecules with an electric dipole moment may act as synchronized sources of electromagnetic radiation, which are capable of creating a destructive interference pattern as a structured region in space, in which the electric field vanishes. Around this region the waves are propagating outwards, but the region itself, with its points and lines of zero field, remains fixed in space. Therefore, the spatial gradient of the electric field amplitude could exercise a force on other biomolecules and steer them to the points of no field. In the example of the salamander, which has to regenerate a hand, it has been shown that the small amount of power the salamander might be capable of putting into radio emission could result in sufficient gradient force to transport neutral molecules with an electric dipole moment to the healing area. On the other hand, the ponderomotive force, which could cause drift of \emph{electrically charged} molecules in the gradient of an oscillating electric field, turned out to be far too small for such an effect. Nevertheless, the results show the principle possibility of a steering mechanism for biological matter based on electromagnetic radiation, which works in a completely different manner compared to the familiar mechanisms of electrophoretic drift and chemical or thermal gradients. One of its distinguishing features is that finely detailed patterns in space can come about far from the sources of the electromagnetic waves. Whether biological systems do make use of such a mechanism for regeneration, wound healing, embryogenesis or just normal growth is not known. It seems that the question of orchestrated electromagnetic emission from living matter in the sub-millimeter to centimeter wavelength regime is only beginning to be explored \cite{JCLin2018}, \cite{UCSanDiego2017}. Therefore, our calculations around the salamander must be understood as a hypothetical example.

In our calculations, the standing pattern in space was realized as minima of the electric field, while most of the emitted radiation went in all other directions without any particular use. This appears as a waste of energy for the organism, and one might wonder, whether the reverse is possible, too. The obvious extension would be a concentration of the electromagnetic energy in the form of a beam with a certain cross section, and the generation of a pattern of minima within this beam. One might also consider the desired pattern to be points of maximum electromagnetic field amplitude, with smaller field amplitudes all around. Although this is the mode of operation of optical tweezers and molecular dipole traps, it seems difficult to realize in an organism. Especially, because the emitting region will always show higher field amplitudes than the target region, and molecules - for which the field would then have to be red-detuned - will not only be attracted to the high field region of the target, but also to the source region.  At any rate, such possibilities must be addressed by further mathematical analysis.

Returning to the situation we have analysed here, an important but open question is, which biomolecules, or which parts in a biological organism, could generate such electromagnetic waves? Microtubules are possible candidates, and although their vibration frequencies might be at the lower end of the interesting spectrum \cite{Cifra2011}, they can be found almost anywhere in an organism. Higher oscillation frequencies might be obtained by their charged C-termini \cite{Priel2005}. But generally, all macromolecules have a broad range of resonance frequencies, and many amino acids have a base frequency in the range of several 10 GHz \cite{Moon2006}. Since proteins usually have an uneven charge distribution in their watery environment, electromagnetic emissions from vibrations are easily conceivable.

Another important and open question is, how the many sources can become synchronized. It is similar to the question of how thousands of muscle fibers can be synchronized into the elegant motion of an elephant's trunk, or how a cuttle fish can become invisible before so many different backgrounds in a matter of seconds \cite{cuttlefish}. While the answers to these questions certainly involve the nervous system, the synchronization of electromagnetic emitters of molecular size will necessitate more basic phenomena like Fr\"ohlich condensation \cite{Lundholm2015}\cite{Nardecchia2018}, or at least a phase locked coupling of the emitters \cite{Damari2019}. Moreover, these emitters need to have a certain distribution of amplitudes. That this may be an extremely stringent requirement can be seen from the fact that in our simulations at most a few thousand emitters were synchronized, while in reality there might have to be orders of magnitude more.

Now, if we assume that some organisms have solved these problems, one might ask, whether the ability of generating electromagnetic radiation with standing patterns in space could also serve for communication between different individuals. One individual could create a field pattern in the volume of another individual, and if the latter has appropriate molecular sensors distributed across its body, it could grasp the message in a single step. And as coarse information of the transmitting individual's position is inadvertently contained in the pattern, the receiving individual could superpose these data with its visual perception, and "see" the message as additional features in the approximate region of the transmitting individual's body. This might be similar to how some birds can "see" the geomagnetic field \cite{Starr2018}. If simple information needs to be communicated to many individuals simultaneously, the created pattern could be a repetitive form all around the transmitting individual. Clearly, such means of communication would work differently from the serial method of our current technologies of electromagnetic information transmission, which relies almost exclusively on patterns in time rather than in space. Of course, temporal modulation of the emitted electromagnetic field is conceivable for organisms, too. In fact, a slow form has been assumed implicitly in our simulations for Fig.2, because the growth zone, and thus the target points for the deposition of molecules, has changed during the process of regeneration.

\section{Conclusion}
In conclusion, we have shown how a large number of randomly distributed sources, all of which emit electromagnetic radiation of the same frequency, can be synchronized to generate a delicately structured zone of vanishing field, which exists as a standing pattern in space. We have argued that, if living organisms were able to create such a field by activating certain molecules with an electric dipole moment, they could use the strong field gradients around such patterns, to attract targeted molecules to this zone in a manner similar to optical tweezers. This could be a novel way of organized molecular transport, as may be needed in embryogenesis, wound healing or regeneration. In the hypothetical example of a salamander having to regrow a hand, we have found that the power expendable by the organism could be sufficient to generate the forces needed for such transport. If not for guiding molecules, the field patterns in space could also serve for communication. The frequencies of the electromagnetic radiation would be in the range of molecular vibrations, which is the GHz to THz regime. Even if future research will not discover such fields from living organisms, generating them artificially may open medical applications.

\section*{Acknowledgment}
I would like to thank Gustav Bernroider for many stimulating discussions on the dynamics of biomolecules and for very helpful remarks on a first version of the manuscript.

\appendix
\section*{Appendices}
\renewcommand{\thesubsection}{\Alph{subsection}}
\renewcommand{\theequation}{A.\arabic{equation}}
\setcounter{equation}{0}

\subsection{Two alternative derivations of the low field zone}
   
It is possible to obtain an intricately shaped zone of very low electric field also with a mathematically different approach than the one followed in section 2. Here we will show two alternative methods.

\subsubsection{\emph{Method 1}}
We begin with the same assumptions about the oscillating dipoles as in section 2, and therefore equations (1)-(3) will be valid again. But in order to determine the unknown amplitude factors $A_j$, $(j=1,...,N-1)$, we try a different boundary condition: We require that the collective electric field shall vanish at time $t=0$ at $N-1$ points in the target region. 
\begin{equation}
\label{Constr1}
E_{tot}(\bold{r}_i,t=0)=0,   \hspace{4mm}\text{for}\hspace{2mm} i=1,...,N-1.
\end{equation}
The rationale behind this approach is that, if the electric field shall disappear at a set of points with nearest-neighbor distances significantly smaller than the wavelength at the same moment in time, it will be very small at any time, because interference minima in space cannot be closer than about half the wavelength. Inserting constraints (\ref{Constr1}) into equation (2) leads to a set of $N-1$ linear equations:
\begin{equation}
\label{GM1}
\begin{pmatrix}
m_{1,1} & m_{1,2} & \cdots & m_{1,N-1} \\
m_{2,1} & m_{2,2} & \cdots & m_{2,N-1} \\
\vdots & \vdots & \ddots & \vdots \\
m_{N-1,1} & m_{N-1,2} & \cdots & m_{N-1,N-1} \\
\end{pmatrix}
\begin{pmatrix}
A_1 \\ A_2 \\ \vdots \\ A_{N-1}
\end{pmatrix}
= -
\begin{pmatrix}
m_{1,N} \\ m_{2,N} \\ \vdots \\m_{N-1,N}
\end{pmatrix}.
\end{equation}
The matrix elements are given by
\begin{equation}
m_{i,j}=f_{ij}\cos\left(2\pi\nu\tau_{ij}-\varphi_j\right),
\end{equation}
for $i=1,...,N-1$ and $j=1,...,N$. The $f_{ij}$ and $\tau_{ij}$ are defined as in eq.(6) and eq.(7), respectively. Solving equations (\ref{GM1}) yields the amplitude factors $A_1$,..., $A_{N-1}$, with which the electric field can be obtained in the general form of eq.(11).

\subsubsection{\emph{Method 2}}

This method is similar to Method 1, but we do not assign arbitrary phases to the oscillating dipoles. Rather, we wish amplitude factor and phase of a dipole to result from the boundary condition, which shall be the same as in Method 1: The electric field shall disappear at the same moment in time at all the target points. Since for each oscillator two parameters will have to be determined, only about half as many oscillators can be uniquely determined from a given number of target points, as compared to Method 1 or to the main method of section 2.

The electric field at a target point $\bold{r}_i$, which results from an oscillating dipole at $\bold{R}_j$ is now more conveniently written as 
\begin{equation}
E_j(\bold{r}_i,t) = A_j f_{ij} \cos\left[2\pi\nu(t-\tau_{ij})\right] 
                  + B_j f_{ij} \sin\left[2\pi\nu(t-\tau_{ij})\right].
\end{equation}
Here, $A_j$ is the amplitude factor of the cosine-component and $B_j$ is the amplitude factor of the sine-component of the electric field. Their ratio is the cotangent of the phase of the oscillator $j$. The factors $f_{ij}$ and the delay times $\tau_{ij}$ are defined as before in eq.(6) and eq.(7), respectively. Let us assume there are $N$ transmitting dipoles. Then their collective electric field at target point $\bold{r}_i$ is:
\begin{equation}
\label{E2}
E_{tot}(\bold{r}_i,t) = \sum_{i=1}^{N} \Big\{ A_j f_{ij}\cos\left[2\pi\nu(t-\tau_{ij})\right] 
                                + B_j f_{ij}\sin\left[2\pi\nu(t-\tau_{ij})\right] \Big\}.
\end{equation}
This equation contains $2N$ free parameters. If we are only interested in the relative field amplitudes between different points and are furthermore not interested in the overall phase, we can assign values to two parameters. We choose $A_N=1$ and $B_N=0$. This leaves $2N-2$ free parameters. In order to determine these parameters we require the electric field to vanish at 
\begin{equation}
M=2N-2
\end{equation} 
points $\mathbf{r}_i$ in the target region simultaneously. The point in time shall again be $t=0$. Then
\begin{equation}
\label{Constr2}
E_{tot}(\bold{r}_i,t=0)=0,  \hspace{4mm} \text{for} \hspace{2mm} i=1,...,M
\end{equation}
Inserting the constraints (\ref{Constr2}) into the equation (\ref{E2}) leads to a set of $M$ linear equations:
\begin{equation}
\label{GM2}
\begin{pmatrix}
p_{1,1} & q_{1,1} & p_{1,2} & q_{1,2} & \cdots & p_{1,N-1} & q_{1,N-1} \\
p_{2,1} & q_{2,1} & p_{2,2} & q_{2,2} & \cdots & p_{2,N-1} & q_{2,N-1} \\
\vdots & \vdots & \vdots & \vdots & \ddots & \vdots & \vdots \\
p_{M,1} & q_{M,1} & p_{M,2} & q_{M,2} & \cdots & p_{M,N-1} & q_{M,N-1} \\
\end{pmatrix}
\begin{pmatrix}
A_1 \\ B_1 \\ A_2 \\ B_2 \\ \vdots \\ A_{N-1} \\ B_{N-1}
\end{pmatrix}
= -
\begin{pmatrix}
p_{1,N} \\ p_{2,N} \\ p_{3,N} \\ p_{4,N} \\ \vdots \\p_{M-1,N} \\ p_{M,N}
\end{pmatrix}
\end{equation}
The matrix elements are defined as
\begin{equation}
p_{i,j}=f_{ij}\cos(2\pi\nu\tau_{ij})
\end{equation}
and
\begin{equation}
q_{i,j}=-f_{ij}\sin(2\pi\nu\tau_{ij}).
\end{equation}
Solving equations (\ref{GM2}) yields the amplitudes $A_1$, $B_1$, ..., $A_{N-1}$, $B_{N-1}$, with which the electric field can be obtained at any point in space, for any time.

It is interesting to note that the results of methods 1 and 2 look practically the same, when displayed in a plot like that of Fig.2, although method 1 allows for arbitrary initial phases. However, in a number of numerical tests with different distributions of source and target points it could be established that the homogeneity in the low field zone as obtained by methods 1 or 2 is not as good as that obtained by the main method presented in section 2.

\subsection{Generation of low field zone with different wavelengths}
\renewcommand\thefigure{B.\arabic{figure}}    
\setcounter{figure}{0} 

For the example with the hand of the salamander in section 2 a fixed wavelength was assumed. Here we give examples of how the same structure appears, if generated with widely different wavelengths but with the same set of source points. Figure B.1 shows the source points and the target points. Figure B.2 shows the generated electric field amplitudes for six different wavelengths.

\begin{figure}[H]
	\centering
	\includegraphics[scale=0.85]{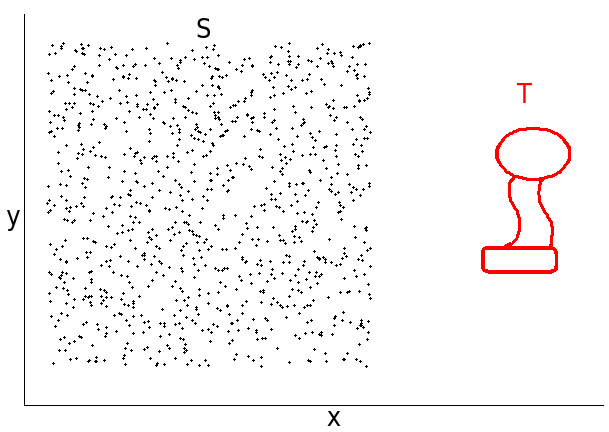}
	\caption{The source points in region S (black crosses) are distributed randomly in the volume of a cuboid with equal extensions in the x- and y-directions, and an extension in the z-direction which is 20 percent of that in x and y. The target points in region T (red) are all in the x-y-plane. They appear as continuous lines, but are actually 1046 separate points with equidistant spacing. The shape has no particular meaning.}
	\label{}
\end{figure}

One can observe in figure B.2, that the basic outline of the target structure can be generated with wavelengths considerably larger than the typical details of the desired structure. But additional unwanted features tend to become more with larger wavelengths, while the contrast (field gradient) becomes weaker. 

\begin{figure}[H]
	\centering
	\includegraphics[scale=0.45]{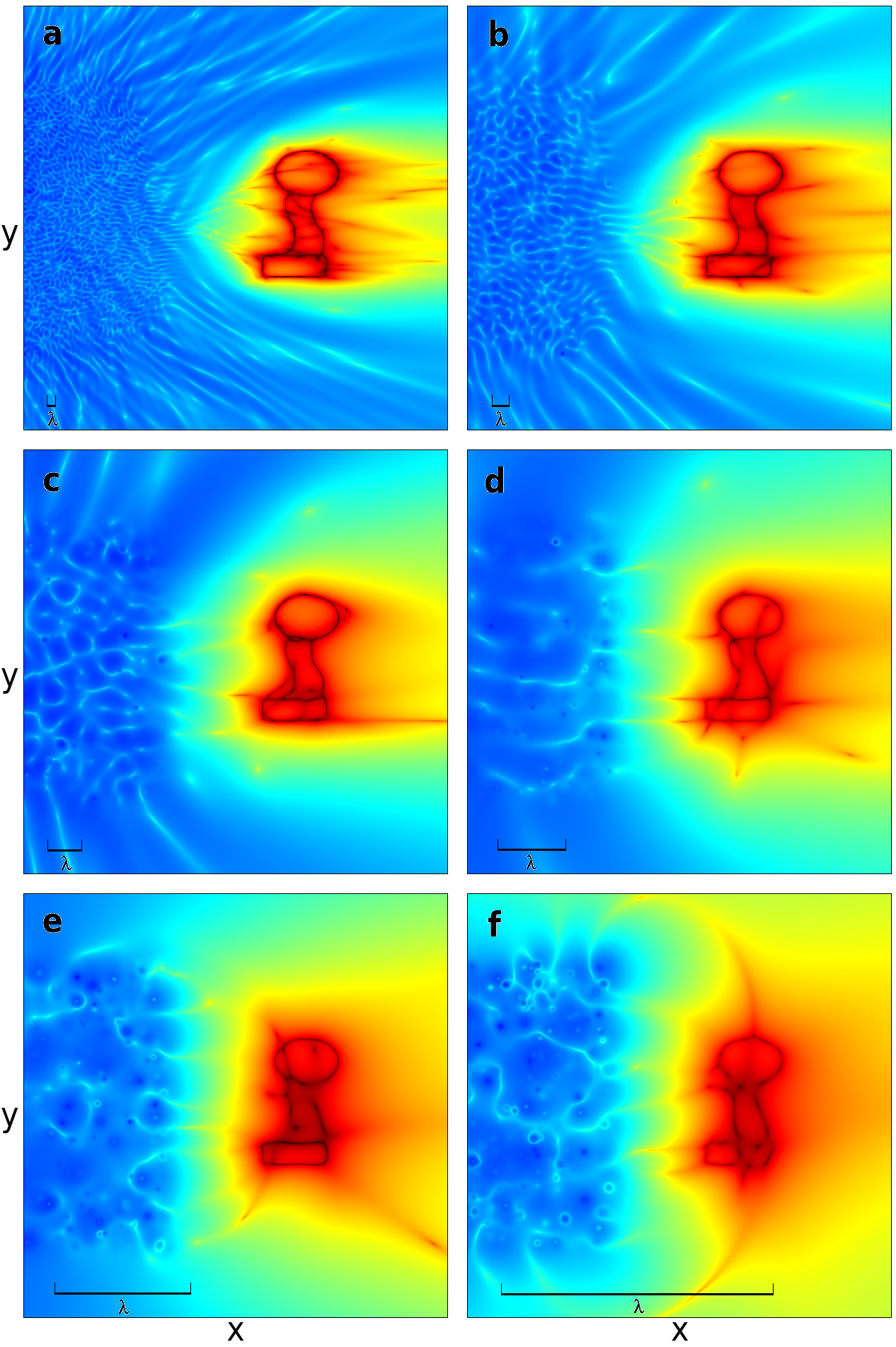}
	\caption{Six examples of the electric field amplitude in the x-y-plane calculated by the method outlined in section 2. Color scale approximately the same as in Fig.2. The wavelength $\lambda$ of the radiation emitted by the source region on the left is smallest in (a), and is doubled from one image to the next.}
	\label{}
\end{figure}

\end{document}